\documentclass[manuscript]{aastex} 
\usepackage{hyperref}
\usepackage{graphicx}
\usepackage{epstopdf}
\usepackage{url}
\usepackage{lscape}
\usepackage{amsmath}

\shorttitle{Anisotropic compact star}
\shortauthors{Pandya \em et al}
\begin{document}
\title{Anisotropic Compact Star Model Satisfying Karmarkar Conditions}
\author{D. M. Pandya} 
\affil{Department of Mathematics, Pandit Deendayal Petroleum University, Raisan, Gandhinagar 382 007, India}
\email{dishantpandya777@gmail.com}
\author{B. Thakore}
\affil{Department of Physics, Pandit Deendayal Petroleum University, Raisan, Gandhinagar 382 007, India}
\email{lightyear1998@gmail.com}
\author{R. B. Goti}
\affil{Department of Physics, Pandit Deendayal Petroleum University, Raisan, Gandhinagar 382 007, India}
\email{rid181198@gmail.com}
\author{J. P. Rank}
\affil{Department of Information and Communication Technology, Pandit Deendayal Petroleum University, Raisan, Gandhinagar 382 007, India}
\email{rankjay0101@gmail.com}
\author{S. Shah}
\affil{Department of Physics, Pandit Deendayal Petroleum University, Raisan, Gandhinagar 382 007, India}
\email{theshlokguy@yahoo.com}
\begin{abstract}
A new class of solutions describing the composition of compact  stars has been proposed, assuming that the fluid distribution inside the star is anisotropic. This is achieved by assuming the appropriate metric potential and then solving Einstein's field equations using Karmarkar conditions [Karmarkar K. R., \textit{Proc. Indian Acad. Sci.} \textbf{27} (1948) 56] to derive the expressions for star density, the radial and tangential pressures in terms of the constants A, B, a paramter `a' and the curvature parameter R. The equations thus obtained have been passed through rigorous conditional analysis. It is further shown that the model is physically viable and mathematically well-behaved, fulfilling the requisite conditions viz., regularity condition, strong energy condition, causality condition, etc. Observed star candidates including EXO 1785-248, SMC X-1, SAXJ1808.43658(SS2), HER X-1, 4U 1538-52, Cen X-3 and LMC X-4 were found to conform to a good approximation through the outcome of this model for a=0.5.

\end{abstract}


\section{\label{sec1}Introduction}
The perusal of interior solutions of Einstein's field equations plays an imperative role in predicting the nature of the star in the final stages of evolution, when the star eventually devolves into an extremely dense, compact astrophysical body. \cite{Ruderman72} and \cite{Canuto74} proposed that the pressure inside highly compact objects shows anisotropy in nature; in essence, it has been categorized into two components: one is radial pressure $p_{r}$ and the other is transverse pressure $p_t$ orthogonal to the former. $ \Delta = p_t - p_r $ is known as an anisotropic factor and $\frac{2\Delta}r $ is an anisotropic force, which can be attractive or repulsive. \cite{Bowers74} documented a paper on the study of the anisotropic distribution of matter, accruing worldwide recognition and acceptance, leading to a major influx of academicians in the field. Anisotropy can be  exposed due to the existence of, but not limited to, solid stellar core or through phase transitions, pion condensation in a star (\cite{Sawyer72}) and the presence of type III-A super fluids (\cite{Sokolov80}). Rotation and electromagnetic fields contribute to the said anisotropy. Through \cite{Herrera97}, the effect of pressure-based anisotropy has been studied in detail. An anisotropic model has been studied by using uniform matter density by \cite{Maharaj89}. The role of the local pressure anisotropy was  studied in detail by \cite{Chan93}. The paper elaborated on how small anisotropies might cause a drastic change in the stability of the system. Some anisotropic compact star models are obtained which admit conformal motion. A new class of interior solutions for anisotropic stars (\cite{Bhar15}) are obtained by  choosing a particular density distribution function of Lorentzian type as showcased by \cite{Nozari09} and \cite{Mehdipour12} which admits conformal motion in higher-dimensional non commutative spacetime. Some researchers(\cite{Lai09}) obtained the anisotropic compact star model  to establish a relation between theory and observations. A model of a relativistic, anisotropic neutron star model at high densities as described by \cite{Heintz75}  employed several simple assumptions and have  shown that for an arbitrary large anisotropy there is no limiting mass for compact stars, albeit the maximum mass of a compact star is still limited to 3-4 $M_{\odot} $. \cite{Sharma01}  assumed a theoretical possibility of generating an anisotropy in strange stars, with densities greater than that of neutron stars but less than that of black holes. \cite{Lai09}  proposed that the anisotropic equations of state are stiffer than the conventional realistic models, i.e., the bag model. Precise solutions corresponding
to statically spherically symmetric anisotropic matter distributions
have been studied and developed by \cite{Bayin82, Krori84, Barreto93, Bondi93,Coley94, Martinez94, PatelMehta95, Singh95,Hernandez99, Bondi99, HarkoMak00, SharmaMukherjee02, HarkoMak02, DevGleiser03, MakHarko03, DevGleiser04, Lake04, BohmerHarko06, Barreto07, BohmerHarko07, Esculpi07, KhadekarTade07, Karmakar07, Abreu07, Herrera08, HerreraOspino08, Ivanov10}.

In the framework of a polytropic model, it is shown that a very low massive compact star can also exist and be  still gravitationally stable even if the polytropic index `n' is greater than three. The properties of neutron stars depend on the assumed description of the matter in their interiors. \cite{Azam16} studied the aforementioned behavior and the physical properties of numerous compact objects. \cite{Alcock86}  and \cite{Haensel86}  proposed a general scheme for compact astrophysical objects which are not composed of neutron matter, however,  the interior density is known to be extremely high. The Randall-Sundrum (RS)  model (\cite{Randall99})  is standing on the concept that our 4-dimensional spacetime is a hypersurface embedded into another 5-dimensional hypersurface. After their work on Brane theory, the study on embedding spacetime drew an ever-increasing number. If an n dimensional space $V$ can be embedded in (n+k)-dimensional space, where k is a minimum number of extra dimensions, then $V_n$ is said to be of embedding class k of n-dimensional space. Two very well-known important solutions, e.g. Schwarzschild's interior solutions and Friedmann universe are of class I, (so, in this case, k=1), on the other hand, Schwarzschild  exterior solution is of class II (k=2) and the Kerr metric (as given by \cite{Kerr63}) is of class $V$(k=5). The Karmarkar  condition relates to class one spacetime. \cite{Pandey81}  presented that the Karmarkar  condition is only a necessary condition for spacetime representing class I. \cite{Sharma13}  described a quadratic equation of state in Finch Skea spacetime, a sub-class of the model described subsequently by \cite{Pandya15}. A comprehensive report of compact stars on pseudo spheroidal spacetime compatible with observational data was provided by \cite{Pandya15b} .

A further requirement has to be  imposed for the sufficiency of the Karmarkar conditions for providing a better insight regarding the topic of the derivation of the \cite{Karmarkar48}. Both uncharged and charged star model of embedding class I spacetime is extensively studied in numerous ways by \cite{Kuchowicz72}, \cite{Bhar16}, \cite{Ratanpal16} and \cite{Pandya17}. 

The paper has been structured as follows:  the basic field equations have  been discussed in Section \ref{sec2}. In Section \ref{sec3} we have given a short discussion about embedding class I spacetime and also obtained a new model. In the next section, we match our interior space-time with the exterior Schwarzschild line element. Whether a model is physically acceptable or not is a question of prime importance in physics, and we have discussed that in Sections \ref{sec5} and \ref{sec6}. The graphs obtained as a consequence have also been described, with a discussion about the model and its further implications being discussed in Section \ref{sec7}. 
\section{\label{sec2} Basic Field Equations}
The interior of static spherically symmetric spacetime in standard co-ordinate $\chi^{a}=(t,r,\theta,\phi) $ is described by the following line element:
\begin{equation}
\label{linel}
ds^{2}=e^{\nu(r)}dt^{2}-e^{\lambda(r)}dr^{2}-r^{2}(d\theta^{2}+\sin^{2}\theta d\phi^{2})
\end{equation}
 where ${\lambda }$ and ${\nu }$ are the functions of the radial co-ordinate $r$. The Einstein field equation is,
\begin{equation}
\label{gijepit} G_{ij}=8\pi T_{ij}
\end{equation}
Here,  $G_{ij} $ is Einstein's tensor, having the form:
\begin{eqnarray}
\label{fieldeqs}
R_{ij}-\frac12Rg_{ij}=G_{ij}
\end{eqnarray}
where $R_{ij}$, $R $ and $g_{ij} $ are the Ricci tensors and Ricci scalar and metric tensor respectively. $T_{ij} $ is the energy-momentum tensor of the underlying fluid distribution. Let us assume that the matter involved in the distribution is anisotropic, by using the general expression, we, therefore, get the expression for energy-momentum tensor as follows:
\begin{equation}
\label{emtens}
T_\xi^\mu =\rho v^\mu v_\xi+p_r\chi_\xi\chi^\mu+p_t(v^\mu v_\xi-\chi_\xi\chi^\mu-g_\xi^\mu)
\end{equation}
 With $v^\mu v_\xi=1=-\chi_\xi\chi^\mu,~\chi_\xi $ is the unit space-like vector and $\upsilon^{\mu}$ is the fluid-4 velocity of the rest frame and therefore $ \upsilon^{\mu} \chi^{\xi}=0 $.  The above formula gives the components of the energy-momentum tensor of an anisotropic fluid at any  point in terms of the density $\rho $, the anisotropic radial and transverse pressures $p_r $ and $p_t $ respectively. With the simple form of a line element, $T_\xi^\mu $ takes the form: 
 \begin{equation}
 \label{tensforms} 
 T_0^{0}=\rho,~ T_1^{1}=-p_r, ~T_2^{2}=T_3^{3}=-p_t\\
 \end{equation}
 And
$T_j^{k}=0 $ if  $j\neq k $. Using (\ref{tensforms}) in (\ref{linel}), we get (\ref{gijepit}) as:
\begin{eqnarray}
\label{rho} 8\pi\rho=\frac{1-\mathrm e^{-{\lambda}}}{\mathrm r^{2}}+\frac{\mathrm e^{-{\lambda}}{\lambda}'}{\mathrm r}\\
\label{radp} 8\pi p_r=\frac{e^{-\lambda}-1}{r^{2}}+\frac{e^{-\lambda}\nu'}r\\
\label{tranp} 8\pi p_t=e^{-\lambda}\left(\frac{\nu''}2+\frac{\nu'^{2}}4-\frac{\nu'\lambda'}4+\frac{\nu'-\lambda'}{2r}\right)
\end{eqnarray}
where differentiation with respect to $r$ is denoted by $'$ and we have chosen G=c=1.   Here G is the gravitational constant and c is the speed of light. The gravitational mass in a sphere of radius $r$ is given by, 
\begin{eqnarray}
\label{masse}
m(r)=4\pi\int_0^{r}\rho(\omega)\omega^{2}d\omega
\end{eqnarray}
 In the following sections, we will, using the field equations as a framework, solve equations (\ref{rho})-(\ref{tranp}) and obtain a physically valid model for compact stars. Based on physical requirements, regularity conditions and stability, we  prescribe bounds on the model parameters, hence our model is compatible with mass and radii of other compact stars. 
\section{\label{sec3} The Model}
In accordance with the aforementioned field equations, it can be seen that we have been given three field equations for five unknown functions:  $e^\nu,~e^\lambda,~\rho,~p_r $, and $p_t $. Hence, to generate an acceptable model of a compact star, we may choose any two of them, and for the model to be a physically realistic model, several physical conditions have to be satisfied by our present model. To generate a particular model of a compact star, let us assume that the co-efficient of $dr^{2} $,  i.e.,  $g_{rr} $ has the following form:
\begin{equation}
\label{elamb}
e^\lambda=1+\frac{r^{2}}{R^{2}(1+a^{2})}
\end{equation}
where `a' is a parameter. Now using the relation between the mass function and the metric potential, we have :
\begin{equation}
\label{elambwm}
e^{-\lambda}=1+\frac{2m(r)}r
\end{equation}
Where m(r) is the mass function of the compact star in equation. The mass function of the star is obtained from equations (\ref{elamb}) and (\ref{elambwm}) as,
\begin{equation}
\label{elambmass}
m(r)=\frac{r^{3}}{2\left(r^{2}+(1+a^{2})R^{2}\right)}
\end{equation}
 The metric potential that we employ in this paper gives a mass function that is monotonically increasing in nature and regular at the center of the compact star [$m(r)_{(r=0)}=0 $]. At the same time, it provides a matter density that gives an acknowledgment of monotonically decreasing nature and gives a finite value at the center of the compact star.  Hence, our chosen metric potential is physically reasonable.

Now, a symmetric tensor $b_{\mu\nu} $ of a 4-dimensional Riemannian space satisfying the Gauss and Codazzi equations is written as :
\begin{eqnarray}
\label{gcod1}
R_{\mu\nu\alpha\beta}=\varepsilon(b_{\mu\alpha}b_{\nu\beta}-b_{\mu\beta}b_{\nu\alpha})\\
\label{gcod2}
b_{\mu\nu;\alpha}-b_{\mu\alpha;\nu}=0~~~~~~~~~~~~~~~~~~~~~~
\end{eqnarray}
can be embedded in 5-dimensional Pseudo-Euclidean space, where (;) represents covariant derivatives and takes the value corresponding to -1 or +1 depending upon the normal to the manifold being time-like or space-like respectively.  Concerning the line element given in (\ref{linel}), the non-zero components of the Riemann curvature tensor can be given as
\begin{eqnarray}
\nonumber
R_{2323}=r^{2}\sin^{2}\theta\lbrack1-e^{-\lambda}\rbrack\\
\nonumber 
R_{1212}=\frac12\lambda^{'}r\\
\nonumber
R_{1224}=0\\
\nonumber
R_{1414}=e^\nu\left[\frac12v^{''}+\frac14v^{'^{2}}-\frac14\lambda^{'}\nu^{'}\right]\\
\nonumber
R_{3434}=\frac r2\sin^{2}\theta v^{'}e^{\nu-\lambda}
 \end{eqnarray}
It can be clearly said that the non-zero components of the symmetry tensor $b_{\mu\nu} $ are $b_{11}, ~b_{22},~ b_{33}, ~b_{44} $ and also $b_{14}(=b_{41}) $ due to its symmetric nature and $b_{33}=b_{22}\sin^{2}\theta $. Upon substituting the components of  $b_{\mu\nu} $ from (\ref{gcod2}), we get:
\begin{equation}
\label{riemann4rank}
R_{1414}=\frac{R_{1212}R_{3434}+R_{1224}R_{1334}}{R_{2323}}
\end{equation}
With $R_{2323}\neq0 $ (\cite{Pandey81}). The space-time that satisfies the condition (equation (\ref{riemann4rank})) corresponds with the space-time of embedding class I.\\
For the condition above, the line element (equation (\ref{linel})) gives the following differential equation:
\begin{equation}
\label{differeqn}
\frac{\lambda'\nu'}{1-e^\lambda}=-2(\nu''+v'^{2})+v'^{2}+\lambda'\nu'
\end{equation}
with $e^\lambda\neq1 $. Solving equation (\ref{differeqn}) we get,
\begin{equation}
\label{enu}
e^\nu=\left(A+B\int\sqrt{e^\lambda-1}dr\right)^{2}
\end{equation} 
Where A and B are constants which are derived after applying necessary conditions. \\
By using equations (\ref{differeqn}) and (\ref{enu}), from equations (\ref{radp}) and (\ref{tranp}), we obtain the pressure anisotropy $\Delta= p_t-p_r $ as,
\begin{equation}
\label{18}
8\pi\Delta=\frac{v^{'}}{4e^\lambda}\left[\frac2r-\frac{\lambda^{'}}{e^\lambda-1}\right]\left[\frac{v^{'}e^\nu}{2rB^{2}}-1\right]
\end{equation}
Once we have the expressions for the metric potential, we can substitute $e^\nu,~ e^\lambda $ into the field equations (\ref{rho})-(\ref{tranp}) to obtain the equations for matter density $\rho $, radial pressure  $p_r $, tangential pressure $p_t $, and the resultant anisotropy $\Delta $ :
\begin{eqnarray}
\label{rhoAB}
\rho=\frac{3 \left(1+a^2\right) R^2+r^2}{8 \pi  \left(\left(1+a^2\right) R^2+r^2\right)^2}\\
\label{radpAB}
p_r=-\frac{2\sqrt{1+a^{2}}AR+B(r^{2}-4(1+a^{2})R^{2})}{8\pi(Br^{2}+2\sqrt{1+a^{2}}AR)\left(r^{2}+(1+a^{2})R^{2}\right)}\\
\label{tranpAB}
p_t=\frac{(1+a^{2})R^{2}\left(-2\sqrt{1+a^{2}}AR+B(r^{2}+4(1+a^{2})R^{2})\right)}{8\pi(Br^{2}+2\sqrt{1+a^{2}}AR)(r^{2}+(1+a^{2})R^{2})^{2}}\\
\label{deltsAB}
\Delta=p_t-p_r=\frac{r^{2}(2\sqrt{1+a^{2}}AR+B(r^{2}-2(1+a^{2})R^{2}))}{8\pi(Br^{2}+2\sqrt{1+a^{2}}AR)(r^{2}+(1+a^{2})R^{2})^{2}}
\end{eqnarray}
Next in Section \ref{sec4}, we shall utilise the boundary conditions for the given metric potential with Schwarzchild's exterior solution to find constants A, B and 'a'. 
\section{\label{sec4}Boundary Conditions}
In this section we match our interior spacetime to the Schwarzchild exterior solution:
\begin{equation}
\label{schwarz}
ds^{2}=\left(1-\frac{2M}r\right)dt^{2}-\left(1-\frac{2M}r\right)^{-1} -r^{2}(d\theta^{2}+\sin^{2}\theta d\phi^{2})
\end{equation}
At the boundary $r=\psi $ ($\psi $ is radius of star), $r$ being the distance from the center of the star to a point inside or on the star's surface, and therefore, it is obvious that $\psi>2M $, where M being mass of dense star considering it as black hole. Now,
\begin{equation}
\label{schwarz1}
1+\frac{r^{2}}{R^{2}(1+a^{2})}=\left(1-\frac{2M}r\right)^{-1}
\end{equation}
Upon considering Schwarzchild exterior metric across the boundary, $r=\psi $:
\begin{equation}
\label{schwarz2}
1+\frac{\psi^{2}}{R^{2}(1+a^{2})}=\left(1-\frac{2M}\psi\right)^{-1}
\end{equation} 
Determining feasible values of geometric parameter R and mass M from equations (\ref{schwarz1}) and (\ref{schwarz2}) in terms of $\psi $, we get:
\begin{eqnarray}
\label{curvepsi}
R=\frac{\psi^{2}+\sqrt{\psi^{4}-16\psi^{2}M^{2}-16a^{2}\psi^{2}M^{2}}}{4(M+a^{2}M)}\\
\label{masspsi}
M=\frac{\psi^{3}}{2(\psi^{2}+(1+a^{2})R^{2})}~~~~~~~~~~~~~~~~~~~~
\end{eqnarray} 
Now, comparing coefficients of $dr{^{2}}$ and $dt{^{2}}$ in Schwarzchild exterior metric and interior spherical spacetime metric,
\begin{eqnarray}
\label{schwarz3}
e^{-\lambda}=\frac{R^{2}(1+a^{2})}{R^{2}(1+a^{2})+\psi^{2}}=\left(1-\frac{2M}\psi\right)\\
\label{schwarz4}
e^\nu=\left(A+\frac{B\psi^{2}}{2\sqrt{(1+a^{2})}R}\right)^{2}
\end{eqnarray}
At $r=0 $, $p_r$ should be equal to 0. Therefore, from equation (\ref{radpAB}), we have:
\begin{equation}
\label{ABrel}
-2\sqrt{1+a^{2}}AR=B(\psi^{2}-4(1+a^{2})R^{2})
\end{equation}
Therefore, from equation (\ref{ABrel}), we get:
\begin{equation}
\label{BinA}
B=\frac{-2A\sqrt{1+a^{2}}R}{\psi^{2}-4(1+a^{2})R^{2}}
\end{equation}
Now, from equations (\ref{schwarz3}) and (\ref{schwarz4}),
\begin{eqnarray}
\label{schwarz5}
A+\frac{B\psi^{2}}{2\sqrt{(1+a^{2})}R}=\sqrt{\frac{R^{2}(1+a^{2})}{R^{2}(1+a^{2})+r^{2}}}
\end{eqnarray}
On substituting the value of B as obtained in equation (\ref{BinA}), we get,
\begin{equation}
\label{Acurvepsi2}
A=\frac{\psi^{2}-4(1+a^{2})R^{2}}{-4(1+a^{2})R^{2}}\sqrt{\frac{R^{2}(1+a^{2})}{R^{2}(1+a^{2})+\psi^{2}}}
\end{equation} 
Solving equations (\ref{BinA}) and (\ref{Acurvepsi2}),
\begin{equation}
\label{Bfinval}
B=\frac1{2\sqrt{R^{2}(1+a^{2})+\psi^{2}}}
\end{equation}

Thus, substituting A and B in equations (\ref{radpAB}), (\ref{tranpAB}) and (\ref{deltsAB}), we get,
\begin{eqnarray}
\label{radppsi}
p_r=\frac{-r^{2}+\psi^{2}}{8\pi(r^{2}+(1+a^{2})R^{2})(r^{2}+4(1+a^{2})R^{2}-\psi^{2})}\\
\label{tranppsi}
p_t=\frac{(1+a^{2})R^{2}(r^{2}+\psi^{2})}{8\pi(r^{2}+(1+a^{2})R^{2})^{2}(r^{2}+4(1+a^{2})R^{2}-\psi^{2})}\\
\label{deltspsi}
\Delta=\frac{r^{2}(r^{2}+2(1+a^{2})R^{2}-\psi^{2})}{8\pi(r^{2}+(1+a^{2})R^{2})^{2}(r^{2}+4(1+a^{2})R^{2}-\psi^{2})}
\end{eqnarray}
\section{\label{sec5}Verification of Physical Parameters}
To verify that the model is physically legitimate, we take into account the conditions set by \cite{Kuchowicz72}, \cite{Buchdahl79}, \cite{Murad15}  and \cite{Knutsen87} :

\subsection{Regularity of metric potential}In our metric, at $r=0, ~e{^{\lambda}}=1$ and $e^{\nu}=A^2$ which are positive constants and, 
\begin{eqnarray}
\label{regsec1con1}
(e^\nu)^{'}=\frac{Br(Br^{2}+2\sqrt{1+a^{2}AR})}{(1+a^{2})R^{2}}\\
\label{regsec1con2}
(e^\lambda)^{'}=\frac{2r}{(1+a^{2})R^{2}}~~~~~~~~~~~~~~~~~~
\end{eqnarray}

Clearly from above equations (\ref{regsec1con1}) and (\ref{regsec1con2}) (derivatives of $e^{v} $ and $e^\lambda $), $(e^{v}{)^{'}}_{(r=0)}=0 $ and $(e^\lambda{)^{'}}_{(r=0)}=0 $ suggest that metric coefficients are regular at $r=0$.

\subsection{Radial Pressure at the Boundary}
The value of $p_r $ should be equal to zero at $r=\psi $. In equation (\ref{radppsi}), the value of $p_r $ at $r=\psi $ is zero. This, along with Fig.\ref{fig2} elucidates the fact that this condition is satisfied.
\subsection{Energy conditions}
\noindent \mbox{\boldmath$(i)~\rho-p_r-2p_t\geq0$} \textbf{(Strong energy condition)}\\
The left hand expression for  strong energy condition is obtained as:
\begin{equation}
\label{stronenercond}
\frac{\sqrt{1+a^2} r^4 R+r^2 \left(3 \left(1+a^2\right)^{3/2} R^3-\sqrt{1+a^2} R \psi ^2\right)-\chi(r)}{4 \pi  \left(\left(1+a^2\right) R^2+r^2\right)^2 \left(\sqrt{1+a^2} r^2 R+\sqrt{1+a^2} R \left(4 \left(1+a^2\right) R^2-\psi ^2\right)\right)}
\end{equation}
Where $\chi(r)=-3 \left(1+a^2\right) R^2 \left(\sqrt{1+a^2} R \psi ^2-2 \left(1+a^2\right)^{3/2} R^3\right)$ . \\
The verification for this condition is being done in Table \ref{tab:1} and Fig. \ref{fig7}.
\newline

\noindent \mbox{\boldmath$(ii)~\rho\geq p_{r},~\rho\geq p_t $} \textbf{(Weak energy condition)}\\
The weak energy indicates that $\rho-p_{r}\geq0$ and $\rho-p_{t}\geq0$.
The equations associated with the weak energy conditions are thus obtained as:
{\small
\begin{eqnarray}
\label{weakenercond1}
\frac{6 a^4 R^4+4 a^2 r^2 R^2+12 a^2 R^4-2 a^2 R^2 \psi ^2+r^4+4 r^2 R^2-r^2 \psi ^2+6 R^4-2 R^2 \psi ^2}{4 \pi  \left(a^2 R^2+r^2+R^2\right)^2 \left(4 a^2 R^2+r^2+4 R^2-\psi ^2\right)}
\end{eqnarray}
and
\begin{eqnarray}
\label{weakenercond2}
\frac{r^2 \left(6 \left(1+a^2\right) R^2-\psi ^2\right)+4 \left(1+a^2\right) R^2 \left(3 \left(1+a^2\right) R^2-\psi ^2\right)+r^4}{8 \pi  \left(\left(1+a^2\right) R^2+r^2\right)^2 \left(4 \left(1+a^2\right) R^2+r^2-\psi ^2\right)}
\end{eqnarray}
} respectively\\
The equations  (\ref{weakenercond1}) and (\ref{weakenercond2}) obtained above satisfy the conditions as elucidated in Table (\ref{tab:1}), since the strong energy condition yields a positive value, indicating that $\rho$ is greater than both $p_{r}$ and $p_t$.

\begin{table}[htbp]
\centering
\scriptsize
\caption{Strong energy condition of various stars at  $r$ = $\psi $ and $r = 0$:}
\label{tab:1}
\vspace{0.01cm}
\begin{tabular}{cccccc}\hline \\
 Star & M & $\psi $ & $ R $ & $(\rho-p_{r}-2p_{t})_{r=0}$ & $(\rho-p_{r}-2p_{t})_{r=\psi }$\\
Name & ($ M_{\odot}$) & (km) & (km) & (MeV fm$^{-3}$) & (MeV fm$^{-3}$)\\ \hline \\
 EXO 1785-248 & 1.3 & 8.84 & 8.99 & 678.58 & 283.77 \\
 SMC X-1 & 1.04 & 8.301 & 9.64 & 641.10 & 305.82  \\
SAX J1808.43658(SS2) & 1.32 & 6.16 & 4.14 & 864.36 & 548.08\\
 Her X-1 & 0.85 & 8.1 & 10.7655 & 542.96 & 294.87 \\
 4U 1538-52 & 0.87 & 7.86 & 10.06 & 613.67 & 321.77 \\	
 CEN X-3 & 1.49 & 9.17 & 8.50 & 695.52 & 267.74 \\
 LMC X-4 & 1.29 & 8.831 & 9.02 & 676.35 & 284.03 \\ \hline
 \end{tabular} 
\end{table}
In Table \ref{tab:1}  we have calculated the values of strong energy condition for various stars at the boundary ($r=\psi$) and at the center ($r=0 $) which is one of the requisite conditions to justify the model's description of a physically realistic stars.
\subsection{Monotone Decrease of Physical Parameters}
The conditions for monotone decrease of physical parameters are as listed below:
 \mbox{\boldmath$\frac{d\rho}{dr}\leq0 $, $\frac{dp_r}{dr}\leq0 $, \textbf{and} $\frac{d p_t}{d r}\leq0$ \textbf{for} $0 \leq r \leq \psi $}.\\
 The respective equations for the parameters mentioned above are obtained as:
\begin{equation}
\label{monotonedec1}
\frac{d\rho}{dr}=-\frac{r \left(5 \left(1+a^2\right) R^2+r^2\right)}{4 \pi  \left(\left(1+a^2\right) R^2+r^2\right)^3}
\end{equation}
\begin{equation}
\label{monotonedec2}
8\pi\frac{dp_r}{dr}=\frac{2(r(r^{4}-4(1+a^{2})^{2}R^{4}-2r^{2}\psi^{2}-4(1+a^{2})R^{2}\psi^{2}+\psi^{4}))}{(r^{2}+(1+a^{2})R^{2})(r^{2}+4(1+a^{2})R^{2}-\psi^{2})^{2}}
\end{equation}
and\\
\begin{equation}
\label{monotonedec3}
8\pi\frac{dp_t}{dr}=\frac{4((1+a^{2})rR^{2}\phi (r)))}{(r^{2}+(1+a^{2})R^{2})^{2}(r^{2}+4(1+a^{2})R^{2}-\psi^{2})^{2}}
\end{equation}
where $\phi (r)$ is,\\$\phi (r)=(-r^{4}+2(1+a^{2})^{2}R^{4}-5(1+a^{2})R^{2}\psi^{2}+\psi^{4}-r^{2}(2(1+a^{2})R^{2}+\psi^{2})$.\\
Due to the complexity of expressions in the right-hand side of equations (\ref{monotonedec2}) and (\ref{monotonedec3}), it is difficult to obtain the sign of the terms in their right-hand side. This becomes clearer in Figs.\ref{fig11} and \ref{fig12} provided below, elucidating the physical ramifications of the equations obtained above, and in the process, satisfying the conditions of monotone decrease.
\subsection{Pressure anisotropy}The difference between radial and tangential pressure should be zero at center of compact star. This condition suggests that at single point the pressure components would be equal. $\Delta_{(r=0)}=0 $ where $\Delta $ is anisotropy of the star. Substituting $r=0$ in equation (\ref{deltspsi}), we find that the condition holds good for this model. This is also verified in Fig. \ref{fig4}.
\subsection{Mass-radius relation }According to \cite{Buchdahl79}, the allowable mass radius relation must satisfy the inequality,                        $\frac {M}{\psi}\leq\frac49 $

We choose the appropriate compact star such that this condition is satisfied, as seen in Table \ref{tab:2}.
\subsection{Redshift}The redshift $z = e^\frac{-\nu}2-1 $ must be a decreasing function of $r$ and finite for  $0\leq z \leq 5$, as given by \cite{BohmerHarko06}:
\begin{equation}
\label{redshift}
z=\frac{4 \sqrt{1+a^2} R \sqrt{\left(1+a^2\right) R^2+\psi ^2}}{4 \left(1+a^2\right) R^2+r^2-\psi ^2}-1
\end{equation}

While the condition cannot be discerned immediately from the equation above, it is satisfied in the graphs, as shown in Fig \ref{fig9}.

In Table \ref{tab:2}, we have given the values of red-shift (which is related to the stability of a relativistic anisotropic stellar configuration), for different realistic stars. The value of the red-shift remains less than 5 as shown.

\subsection{Stability Conditions}
\mbox{\boldmath$(i)~ 0\leq (\frac{dp_{r}}{d\rho}) \leq 1, 0 \leq (\frac{dp_{t}}{d\rho})\leq 1$ for $ 0\leq r\leq\psi $} \textbf{(Causality condition)}\\
The values for the radial speed of sound waves $ (\frac{dp_{r}}{d\rho}) $ (denoted by $\upsilon_r^{2}$) and transverse speed of sound waves $ (\frac{dp_{t}}{d\rho}) $ (denoted by $\upsilon_t^{2}$) between $r=0$ and $ r=\psi $ for different stars have been calculated exhaustively and mentioned in Table \ref{tab:3} , and found to comply with the requirement. These velocities are higher in magnitude in more compact stars, the expressions for which are calculated in Subsection \ref{variation}\\
\indent\mbox{\boldmath$(ii)~\Gamma_{r} = \frac{\rho+p_r{\displaystyle}}{p_r}\frac{dp_r}{d\rho} $ \textbf{(Relativistic adiabatic index)} } \\
For a relativistically stable stellar model, the adiabatic index stated must be greater than  $ 1.3333... $ in the prescribed range $ 0 \leq  r \leq  \psi $.  The conditional testing and validation is shown in Table \ref{tab:2}, and Fig.\ref{fig8}.
\begin{table}[ht]
	\centering
	\scriptsize
	\caption{Red-shift, adiabatic index and Buchdahl Ratio for  various stars at $r = \psi $ and $r = 0$:}
	\label{tab:2}
	\vspace{0.01cm}
	\begin{tabular}{ccccccc}\hline \\
		Star & M & $\psi $ & $ R $ & $z _{(r=\psi)}$ & $\Gamma_{r (r=0)}$ & $\frac{M}{\psi}$\\
		Name & ($ M_{\odot}$) & (km) & (km) & Red-shift & Adiabatic index & Buchdahl Ratio\\ \hline \\
		EXO 1785-248 & 1.3 & 8.84 & 8.99 & 0.332 & 1.686 &0.21\\
		SMC X-1 & 1.04 & 8.301 & 9.64 & 0.262 & 1.891  &0.18\\
		SAX J1808.43658(SS2) & 1.32 & 6.16 & 4.14 & 0.665 & 1.528&0.32\\
		Her X-1 & 0.85 & 8.1 & 10.7655 & 0.205 & 2.183 &0.15\\
		4U 1538-52 & 0.87 & 7.86 & 10.06 & 0.219 & 2.092& 0.16\\	
		CEN X-3 & 1.49 & 9.17 & 8.50 & 0.390 & 1.585 &0.24\\
		LMC X-4 & 1.29 & 8.831 & 9.02 & 0.329 & 1.692 &0.21\\ \hline
	\end{tabular} 
\end{table}
\subsection{Variation of Physical Parameters \label{variation}}The  variation of density $\rho$ with respect to the radial variable $r$ is given as shown in equation (\ref{monotonedec1}). Since $\frac{d\rho}{dr} \leq0$, for $0 \leq r\leq\psi $, the density distribution decreases radially outward. To fulfill Herrera's stability condition (\cite{Herrera92}), we calculate the radial and transverse velocity expressions as
\begin{eqnarray}
\label{varphys1}
v_{r}^2=\frac{\partial p_r}{\partial\rho}=-\frac{\left(a^2R^2+r^2+R^2\right)\left(-4 a^4 R^4-8a^2R^4-4a^2R^2 \psi ^2+r^4-2r^2 \psi ^2-4 R^4-4 R^2 \psi ^2+\psi^4\right)}{\left(5a^2R^2+r^2+5 R^2\right)\left(4a^2R^2+r^2+4R^2-\psi^2\right)^2}\\
\label{varphys2}
v_{t}^2=\frac{\partial p_t}{\partial\rho}=-\frac{2 \left(1+a^2\right) R^2 \cdot \Omega(r)}{\left(5 a^2 R^2+r^2+5 R^2\right) \left(4 a^2 R^2+r^2+4 R^2-\psi ^2\right)^2}~~~~~~~~~~~~~~~~~~~~~~~~~~~~~~~~~~~~~~~~~~~~~~~~
\end{eqnarray}
where \\ $\Omega(r)=\left(2 a^4 R^4-2 a^2 r^2 R^2+4 a^2 R^4-5 a^2 R^2 \psi ^2-r^4-2 r^2 R^2-r^2 \psi ^2+2 R^4-5 R^2 \psi ^2+\psi ^4\right)$.

The radial and transverse speeds of sound waves $\upsilon_r $ and $\upsilon_t $ are plotted against $r$ in Fig.\ref{fig5} and \ref{fig6} respectively. 

\begin{table}[ht]
\centering
\scriptsize
\caption{Radial and Transverse velocities of various stars at $r = \psi $ and $r = 0$ :}
\label{tab:3}
\vspace{0.01cm}
\begin{tabular}{cccccccccc}\hline \\
 Star & M & $\psi $ & $ R $ & $(\frac{dp_{r}}{d\rho})_{r=0}$ & $(\frac{dp_{t}}{d\rho})_{r=0}$&$\left(v_{t}^{2}-v_{r}^{2}\right)_{r=0}$ &$(\frac{dp_{r}}{d\rho})_{r=\psi}$ & $(\frac{dp_{t}}{d\rho})_{r=\psi}$&$\left(v_{t}^{2}-v_{r}^{2}\right)_{r=\psi}$\\
Name & ($ M_{\odot}$) & (km) & (km) &  & & & & &\\ \hline \\
 EXO 1785-248 & 1.3 & 8.84 & 8.99 & 0.125 & 0.049 &  -0.076 &0.136 &  0.086& -0.050\\
 SMC X-1 & 1.04 & 8.301 & 9.64 & 0.104 & 0.021 &   -0.083&0.113 &  0.056& -0.057\\
SAX J1808.43658(SS2) & 1.32 & 6.16 & 4.14 & 0.319 &  0.299  & -0.02   & 0.283 &  0.249&-0.034\\
 Her X-1 & 0.85 & 8.1 & 10.7655 & 0.089 & 0.002 &   -0.087& 0.097 & 0.032&-0.065\\
 4U 1538-52 & 0.87 & 7.86 & 10.06 & 0.093 & 0.007 &   -0.086&0.101 &  0.038& -0.063\\	
 CEN X-3 & 1.49 & 9.17 & 8.50 & 0.146 & 0.076 &    -0.070&0.157 &   0.113&-0.044\\
 LMC X-4 & 1.29 & 8.831 & 9.02 & 0.124 & 0.048 &    -0.076 &0.135 &  0.086&-0.049\\ \hline
 \end{tabular} 
\end{table}
\section{\label{sec6}Physical Analysis}
In order to examine the compatibility of the model with observational data, we assumed the metric potential in the form $e^{\lambda}=1+\frac{r^{2}}{R^{2}(1+a^2)}$, where $a=0.5$; and considered known strange star candidates viz.,  EXO 1785-248, SMC X-1, SAX J1808, Her X-1, 4U 1538, Cen X-3, LMC X-4 (\cite{Gangopadhyay13} and \cite{ozel09}). By choosing the mass M and radius $\psi$ of the respective star, using equation (\ref{masspsi}), the value of the corresponding geometric (curvature) parameter R is obtained in Table \ref{tab:4}, along with the corresponding central and surface densities. Considering speed of light, c to be 299792 km s$^{-1}$ and value of universal Gravitational constant, G to be $6.67430\cdot 10^{-20}$ km$^{-3}$ kg$^{-1}$  s$^{-2}$.
\begin{table}[ht]
\centering
\scriptsize
\caption{Central and Surface densities of various stars are shown in the following table:}
\label{tab:4}
\vspace{0.01cm}
\begin{tabular}{cccccc}\hline \\
 Star & M & $ \psi $ & $ R $ & $\rho_c$ & $ \rho_s$\\
Name & ($ M_{\odot} $) & (km) & (km) & (MeV fm$^{-3}$) & (MeV fm$^{-3}$)\\ \hline \\
 EXO 1785-248 & 1.3 & 8.84 & 8.99 & 925.59 & 361.85 \\
 SMC X-1 & 1.04 & 8.301 & 9.64 & 1051.84 & 411.23  \\
SAX J1808.43658(SS2) & 1.32 & 6.16 & 4.14 & 1910.07 & 746.735\\
 Her X-1 & 0.85 & 8.1 & 10.7655 & 1104.69 & 431.875 \\
 4U 1538-52 & 0.87 & 7.86 & 10.06 & 1171.39 & 457.952 \\	
 CEN X-3 & 1.49 & 9.17 & 8.50 & 1121.24 & 438.84 \\
 LMC X-4 & 1.29 & 8.831 & 9.02 & 860.428 & 336.381 \\ \hline
 \end{tabular} 
\end{table}

\begin{figure}
\includegraphics[scale=1.3]{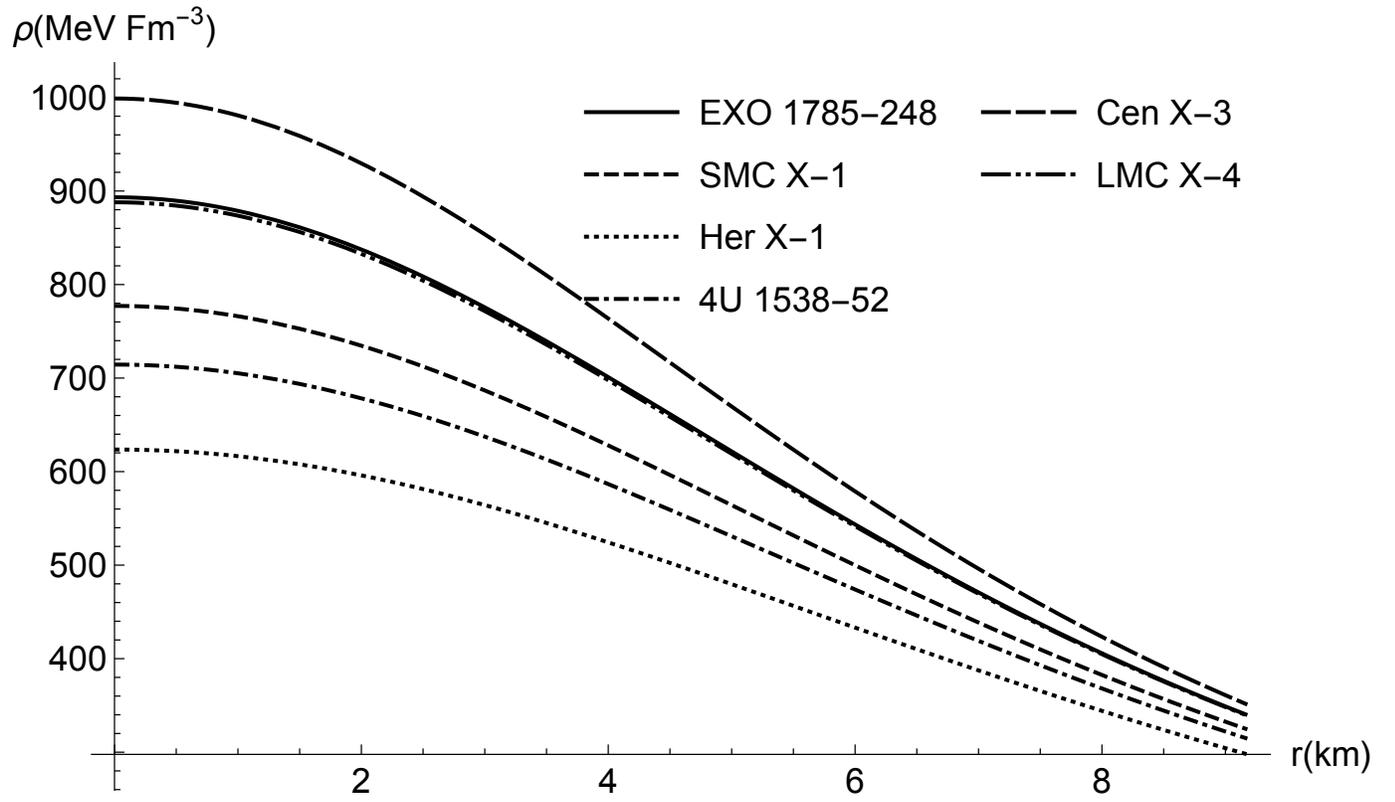}
\caption{Density profile \label{fig1}}
\end{figure}
\begin{figure}
\includegraphics[scale=1.2]{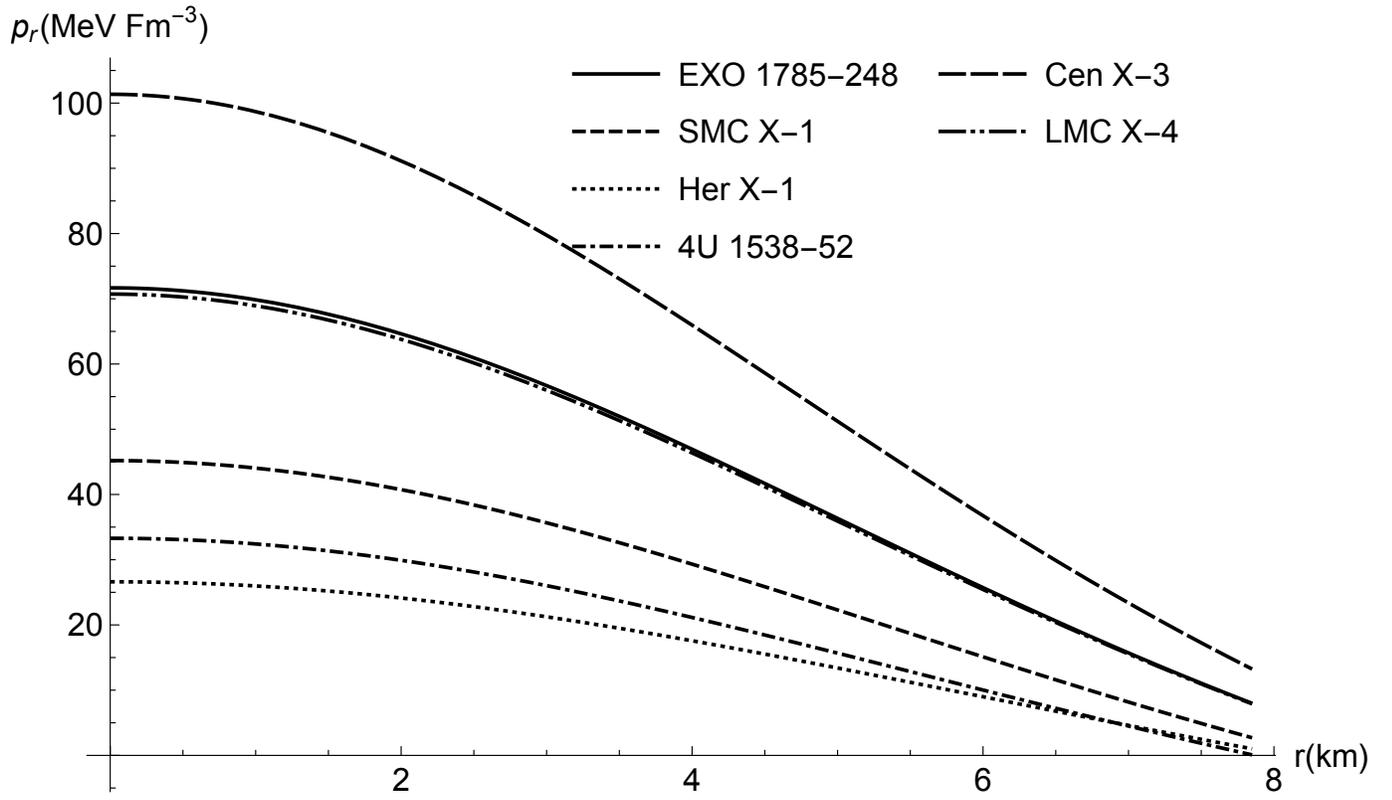}
\caption{Radial pressure profile \label{fig2}}
\end{figure}
\begin{figure}
\includegraphics[scale=1.15]{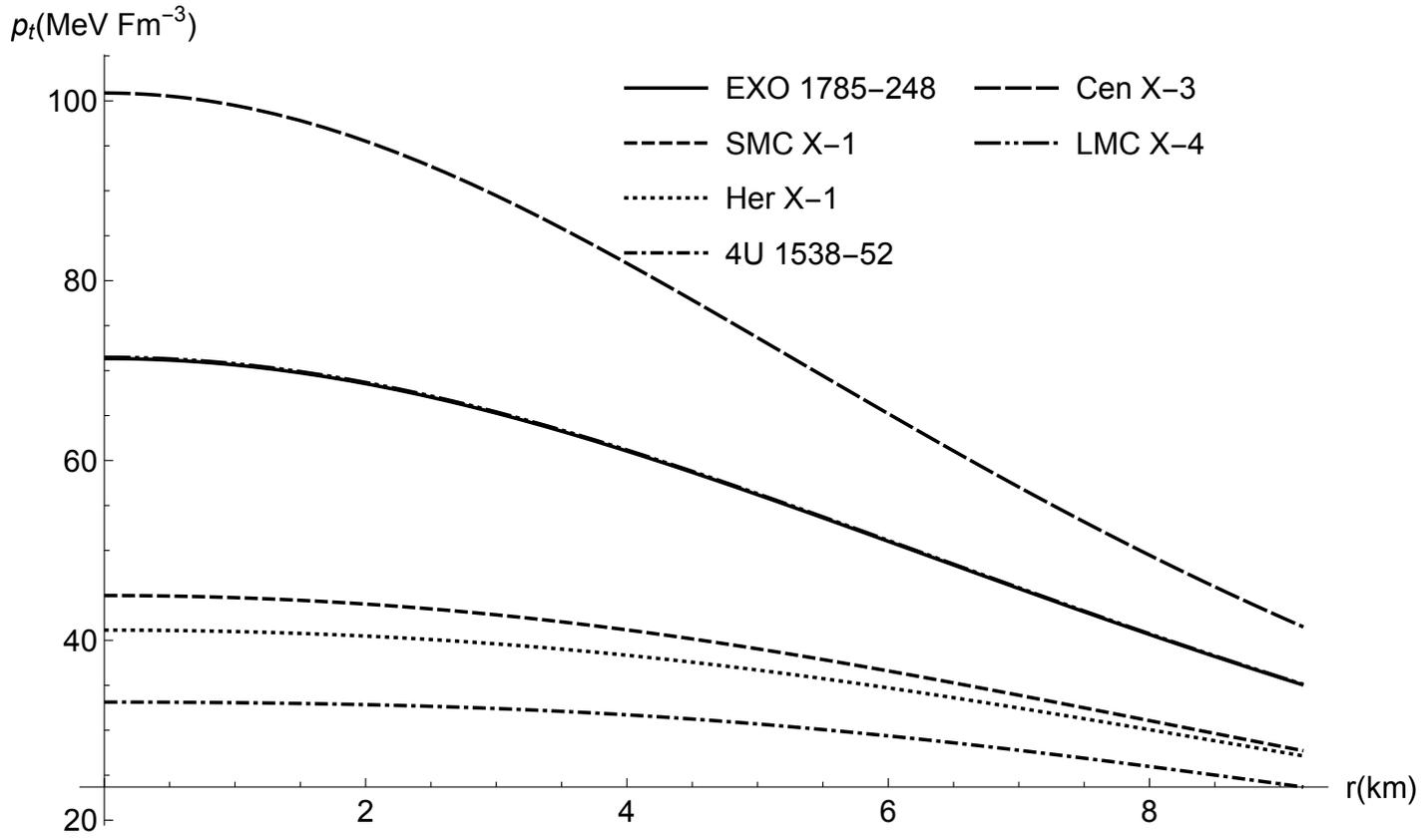}
\caption{Transverse pressure profile \label{fig3}}
\end{figure}
\begin{figure}
\includegraphics[scale=1.2]{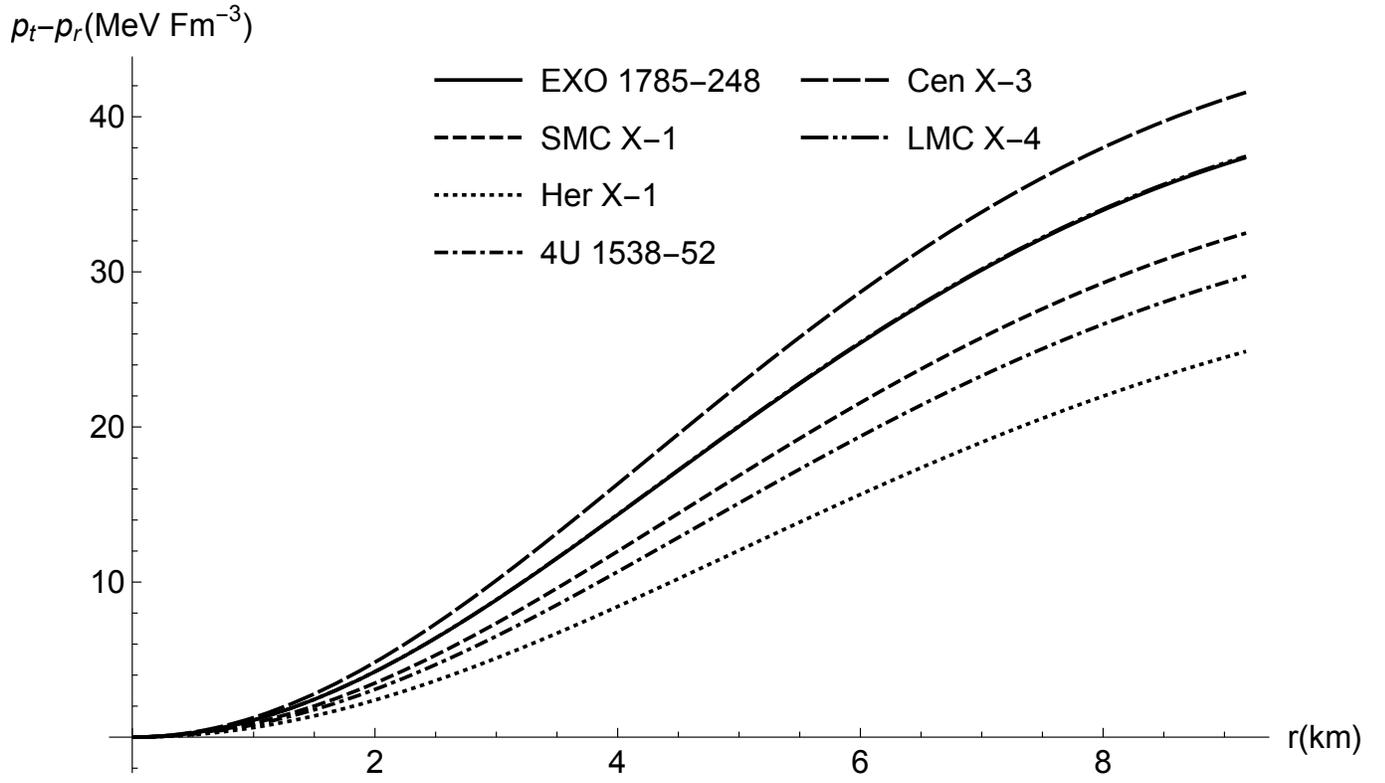}
\caption{Anisotropy profile \label{fig4}}
\end{figure}
\begin{figure}
\includegraphics[scale=1.1]{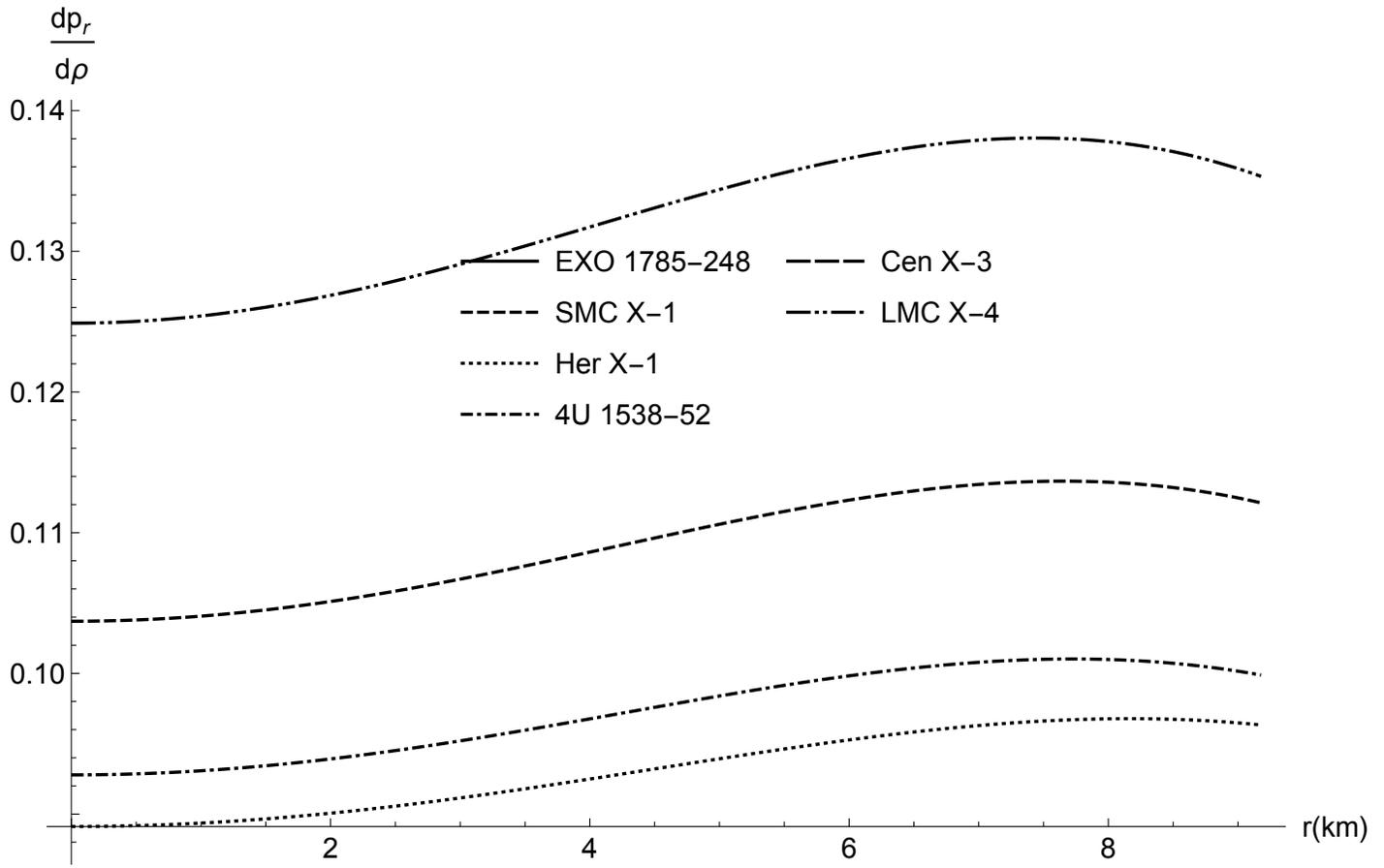}
\caption{Radial sound speed profile \label{fig5}}
\end{figure}
\begin{figure}
\includegraphics[scale=0.9]{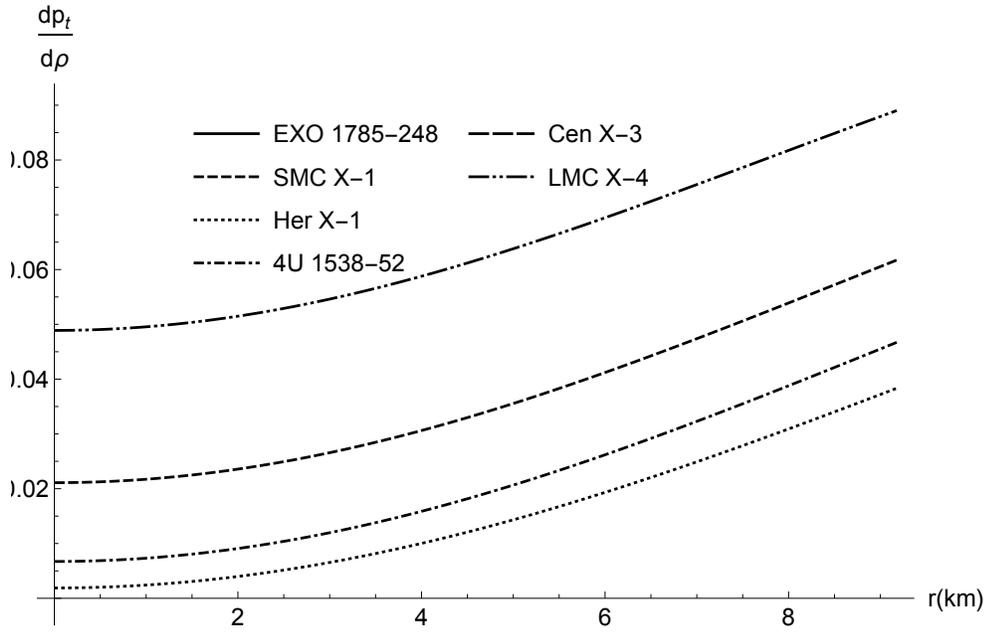}
\caption{Transverse sound speed profile \label{fig6}}
\end{figure}
\begin{figure}
\includegraphics[scale=1.15]{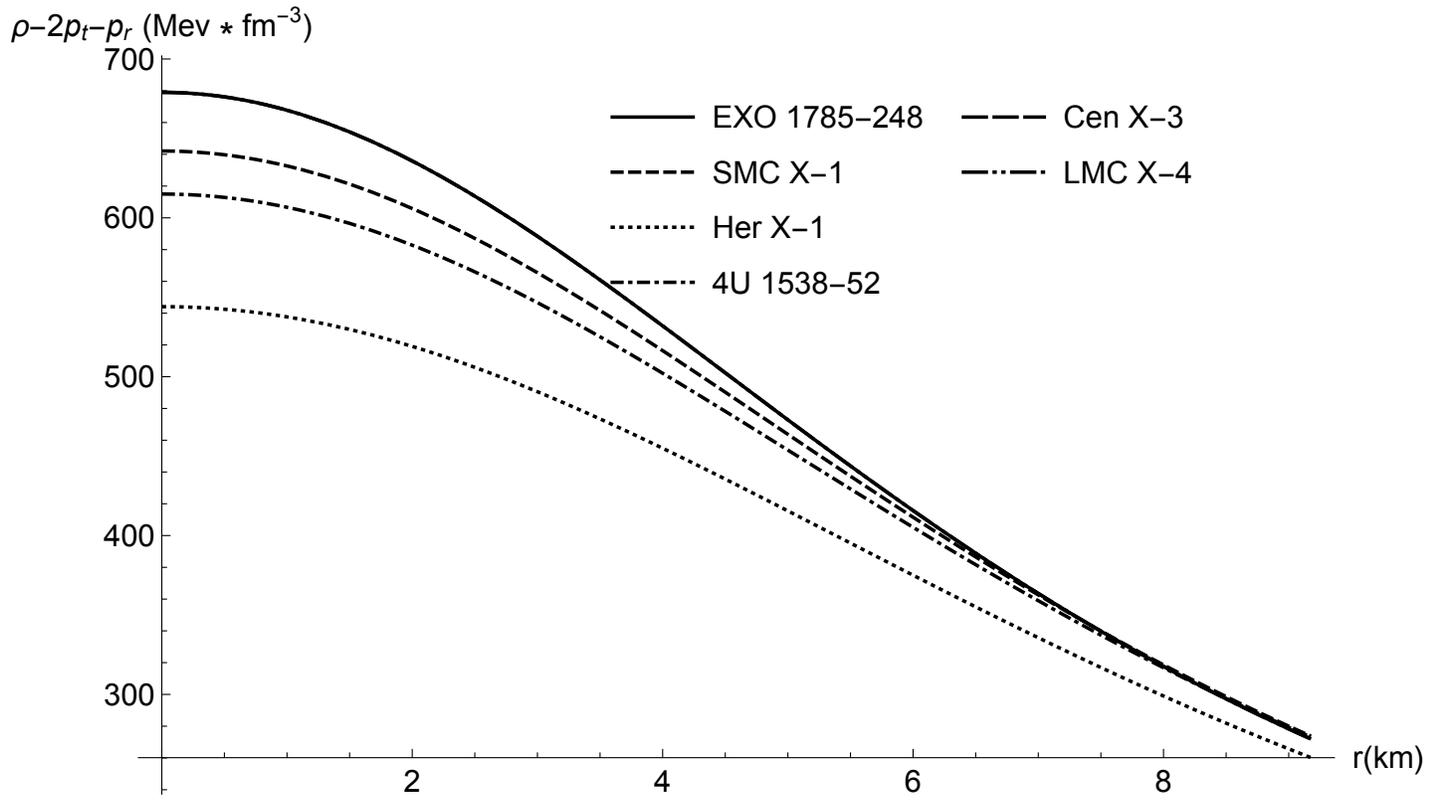}
 \caption{Strong Energy Condition profile \label{fig7}}
 \end{figure}
\begin{figure}
\includegraphics[scale=1.65]{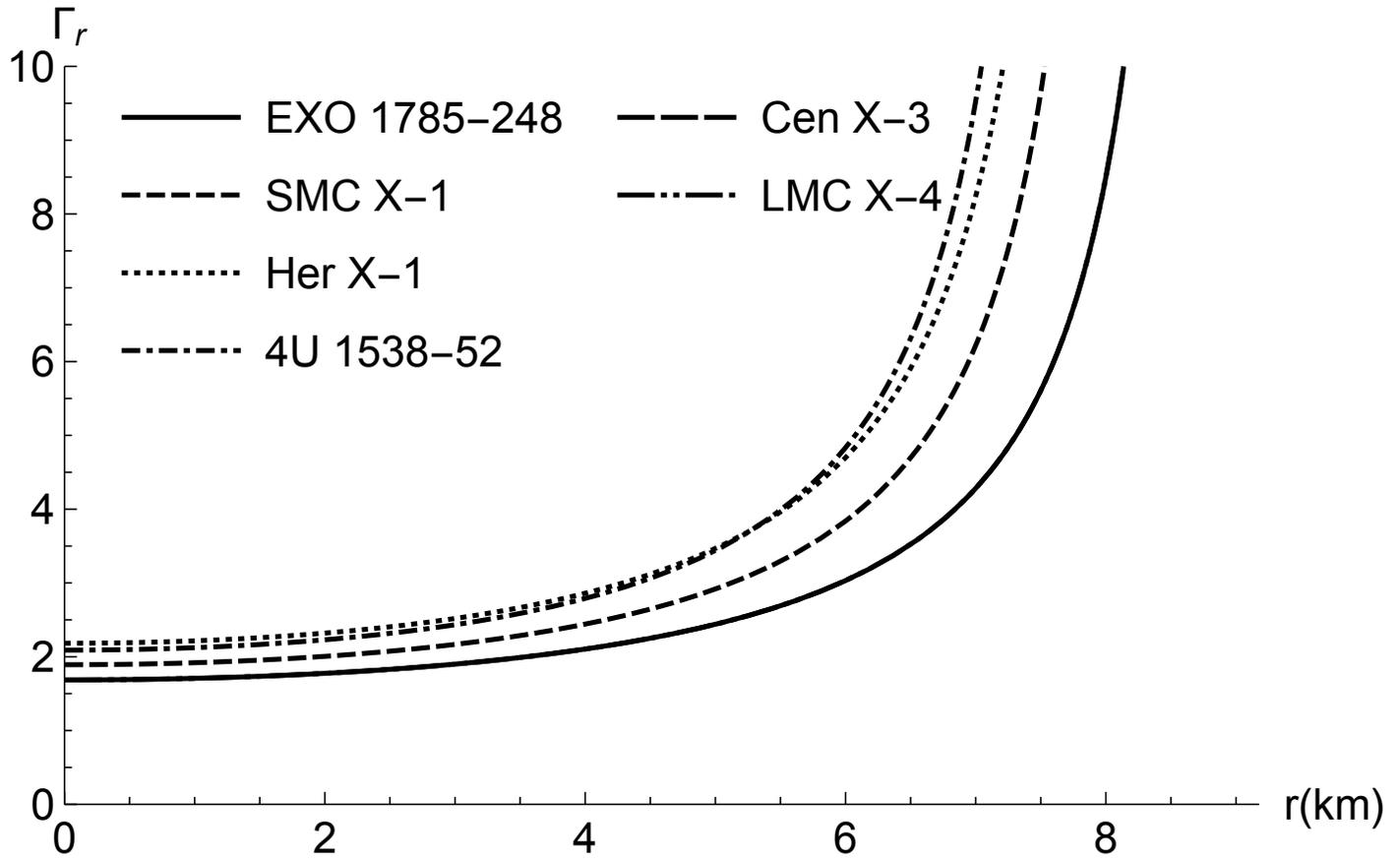}
\caption{Adiabatic Index \label{fig8}}
\end{figure}
\begin{figure}
\includegraphics[scale=1.15]{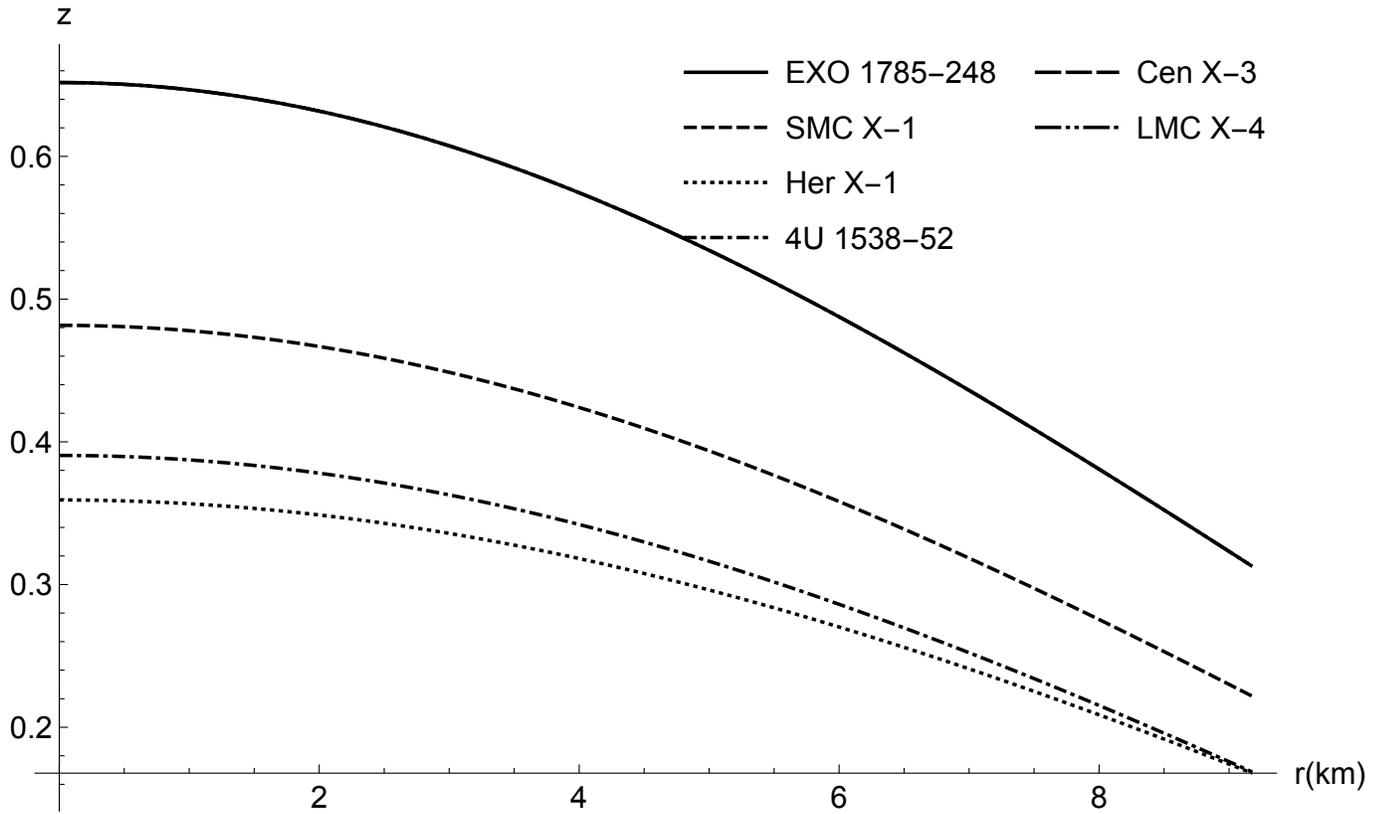}
\caption{Gravitational Redshift. \label{fig9}}
\end{figure}

Using values of $\psi $ and $R$ from Table \ref{tab:4}, variations of physical parameters of different stars are plotted, with $\rho$, $p_{r}$ and $p_t$ having a unit of MeV fm$^-3$. In Fig.\ref{fig1}, \ref{fig2} and \ref{fig3}, we have shown variation of the matter density $\rho $, radial pressure $p_r$ and transverse pressure $p_t$  respectively against radial parameter $r$, all of which are monotonically decreasing functions.

The measure of anisotropy, $\Delta$ is plotted in Fig.\ref{fig4}. We can see that for every individual star, value of  $\Delta>0 $ for $0<r\leq\psi $ . The positive nature of $\Delta$ offers a repulsive force which helps to hold the model against the collapse due to gravity. Also, at $r = 0$,  $\Delta = 0$, which is necessary to construct a well-behaved compact star model.  Anisotropy increases as we move radially outward. Figs. \ref{fig5} and \ref{fig6} illustrate the condition that radial and transverse sound speed profiles lie in [$0, 1$]. Fig.\ref{fig7}  shows that the strong energy condition satisfies for all stars throughout the distribution. The adiabatic index $\Gamma_{r}$ value is plotted in Fig.\ref{fig8}.  One can observe that the value greater than $\frac{4}{3}$  is achieved for all stars throughout the distribution.  As the value of $\Gamma_{r}$ decreases,  stars become less stable as their compactness increases. In Fig.\ref{fig9}, variation of red-shift $z=\sqrt{e^{-\nu(r)}}-1 $  is plotted, which is a decreasing function of $r$. Stars with more compactness demosntrates more redshift.

\begin{figure}
\includegraphics[scale=1.6]{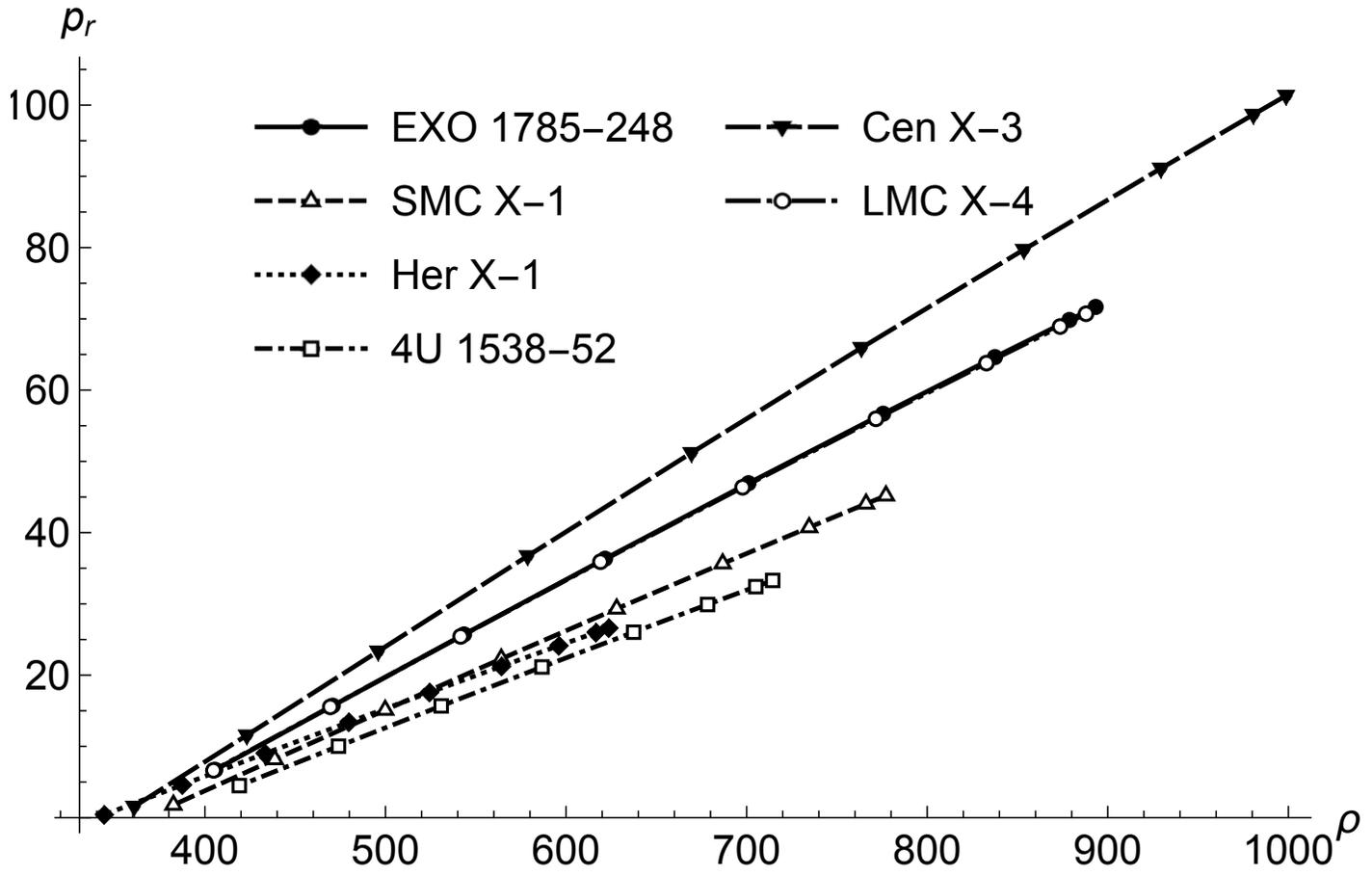}
\caption{Equation of State. \label{fig10}}
\end{figure}

\begin{figure}
\includegraphics[scale=1]{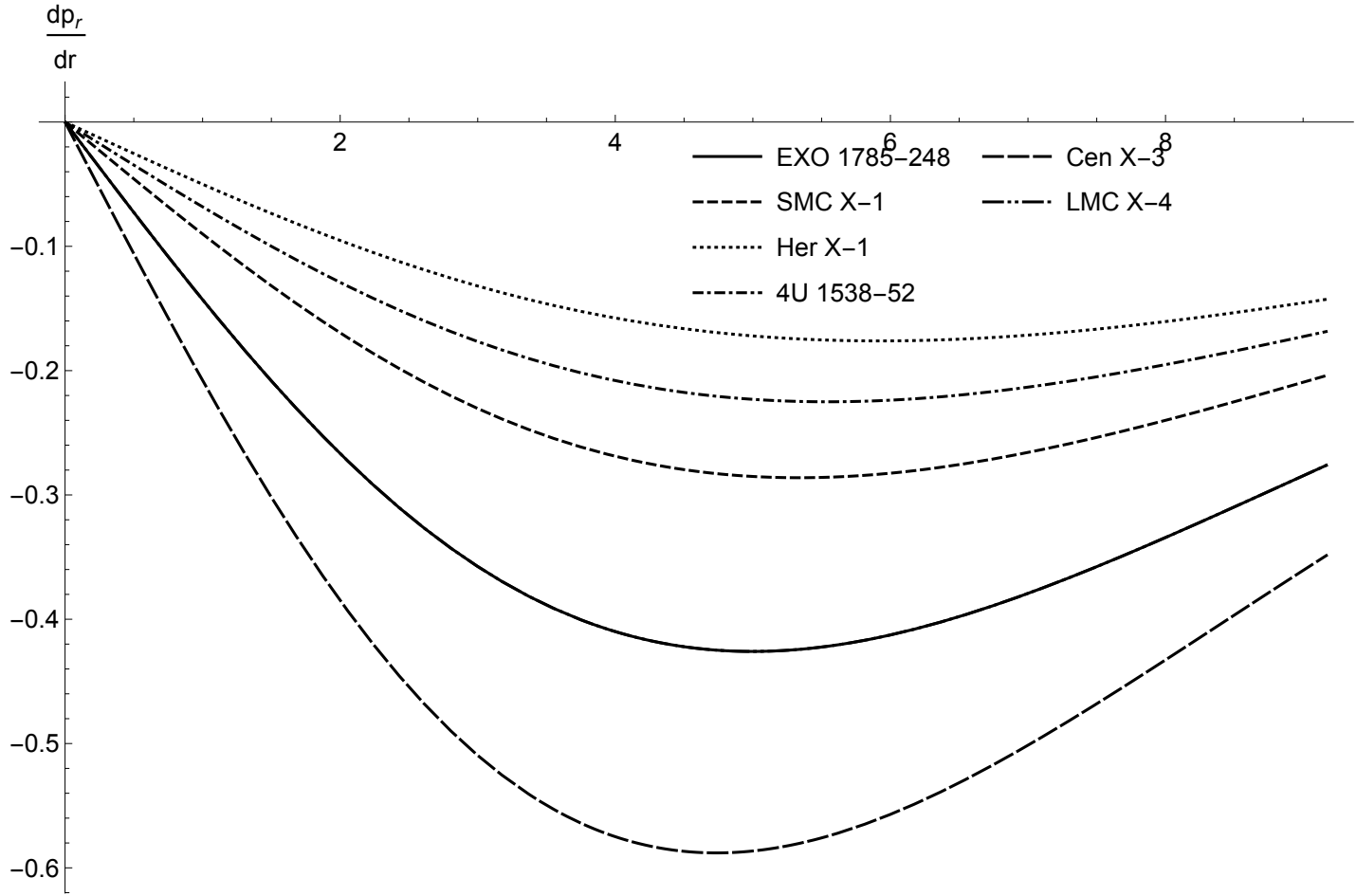}
\caption{Radial Pressure Gradient. \label{fig11}}
\end{figure}

\begin{figure}
\includegraphics[scale=0.95]{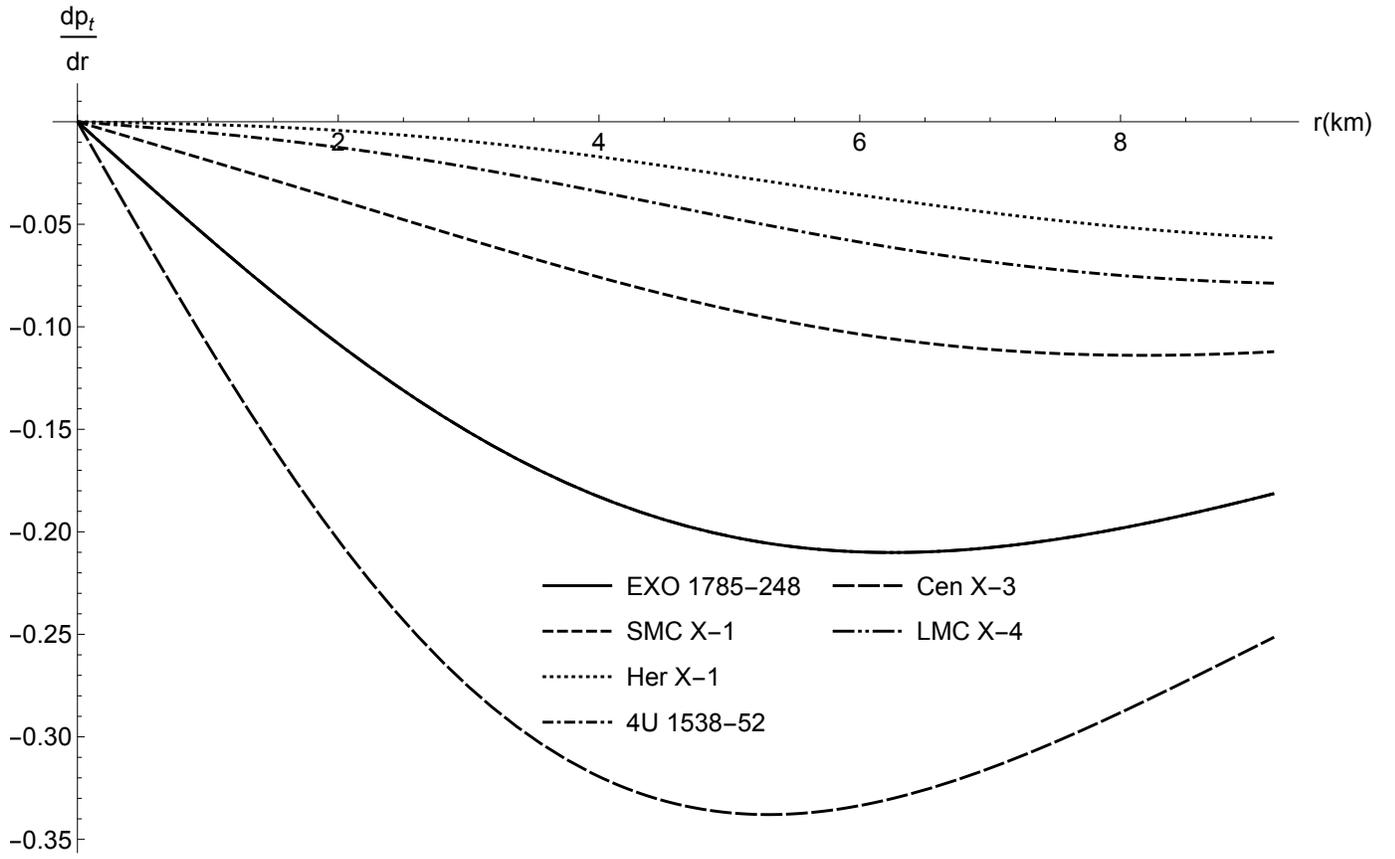}
\caption{Transverse Pressure Gradient. \label{fig12}}
\end{figure}

Fig.\ref{fig10} represents the Equation Of State (EOS) for all the stars. While the formulation of an EOS is essentially governed by the physical laws of the system, parametric relations of the energy-density and the radial pressure from the mathematical model may be useful in predicting the composition of the system. The EOS here  exhibits linear behaviour, which implies that radial pressure and density expand at a uniform proportion throughout the star. An intriguing characteristic of this class of solutions is that the EOS can be applied to strange stars with quark matter. Another salient feature of the model is that the EOS was found to be linear without apriori assumptions of a linear nature.\\
\section{\label{sec7}Discussion}
Based on the aforementioned Karmarkar conditions, we have generated closed form solutions of Einstein's field equations representing an ansiotropic analogue of static spherically symmetric space-time. The model successfully satisfies all the requisite physical and mathematical parameters, thus providing a legitimate structural model described by fixing a suitable  value of  paramter `a', the curvature paramter ($R$), the radius of the star ($\psi$) and the mass of the star ($M$). The model describes multiple star candidates of various masses and radii up to a reasonable approximation. It is found that stars whose compactness is more accommodate more density, pressure and $\Delta$. The red-shift increases with compactness while the value of $\Gamma_{r}$ decreases with compactness showing that the stability decreases with increase in compactness. The internal and external pressures are described as collective forces, arising from neutron-hyperon interactions, phase transitions, electromagnetic interactions, etc. The current paper can be utilised to map the collective quark-gluon interactions at the center of such strange stars, which may provide crucial information about the star's inherent anisotropy. 



\begin{thebibliography}{00}
\bibitem[\protect \citeauthoryear{R.F Sawyer}{1972}]{Sawyer72} Sawyer R.F, {\it Phys.Rev.Lett.} {\bf29(6)} (1972) 382.
\bibitem[\protect \citeauthoryear{Ruderman}{1972}]{Ruderman72} Ruderman R., {\it Astro. Astrophys.} {\bf10} (1972) 427.
\bibitem[\protect \citeauthoryear{Canuto}{1974}]{Canuto74} Canuto V., {\it Annu. Rev. Astron. Astrophys.} {\bf12} (1974) 167.
\bibitem[\protect \citeauthoryear{Bowers and Liang}{1974}]{Bowers74} Bowers R. and Liang E., {\it Astrophys. J.} {\bf188} (1974) 657.
\bibitem[\protect \citeauthoryear{Maharaj and Marteens}{1989}]{Maharaj89} Maharaj S. D. and  Marteens R., {\it Gen. Relativ. Grav.} {\bf21} (1989) 899.
\bibitem[\protect \citeauthoryear{A.I Sokolov}{1980}]{Sokolov80} Sokolov A.I, {\it Sov. Phys. JETP} {\bf52(4)} (1980) 575.
\bibitem[\protect \citeauthoryear{Herrera and Santos}{1997}]{Herrera97} Herrera L. and Santos N. O., {\it Phys. Rep} {\bf286} (1997) 53.
\bibitem[\protect \citeauthoryear{Bhar{\em et al}}{2015}]{Bhar15} Bhar P.,Rahaman F.,Ray S.,Chatterjee V. {\it Eur.Phys.J.C} {\bf75} (2015) 190.
\bibitem[\protect \citeauthoryear{Nozari}{2009}]{Nozari09} Nozari K.,Mehdipour S.H, {\it JHEP} {\bf0903} (2009) 61.
\bibitem[\protect \citeauthoryear{Mehdipour}{2012}]{Mehdipour12} Mehdipour S.H, {\it Eur.Phys.J.C} {\bf64} (2012) 80.
\bibitem[\protect \citeauthoryear{Lai and Xu}{2009}]{Lai09} Lai X. Y., Xu R. X., {\it Astropart. Phys} {\bf31} (2009) 128.
\bibitem[\protect \citeauthoryear{Heintzmann and Hillebrandt}{1975}]{Heintz75} Heintzmann. and Hillebrandt, {\it Astron. Astrophys.} {\bf38} (1975) 51.
\bibitem[\protect \citeauthoryear{Azam {\em et al}}{2016}]{Azam16} Azam M{\em et al}, {\it Chin.Phys.Lett.} {\bf33} (2016) 070401.
\bibitem[\protect \citeauthoryear{Alcock {\em et al}}{1986}]{Alcock86} Alcock C., {\it Astrophys. J} {\bf310} (1986) 261.
\bibitem[\protect \citeauthoryear{Haensel {\em et al}}{1986}]{Haensel86} Haensel P et al., {\it Astron. Astrophys.} {\bf160} (1986) 121.
\bibitem[\protect \citeauthoryear{Randall and Sundrum}{1999}]{Randall99} Randall L. and Sundrum R., {\it Phys.Rev.Lett.} {\bf83} (1999) 3370.
\bibitem[\protect \citeauthoryear{R.P Kerr}{1963}]{Kerr63} Kerr R.P, {\it Phys. Rev.Lett.} {\bf11} (1963)237.
\bibitem[\protect \citeauthoryear{Pandey and Sharma}{1981}]{Pandey81} Pandey S.N and Sharma S.P, {\it Gen. Relativ. Gravit.} {\bf14} (1981) 113.
\bibitem[\protect \citeauthoryear{Sharma and Ratanpal}{2013}]{Sharma13} Sharma R. and Ratanpal B.S, {\it Int.J.Mod.Phys.D} {\bf356(2)} (2013) 1350074.
\bibitem[\protect \citeauthoryear{Pandya {\em et al}}{2015}]{Pandya15} Pandya D.M,Thomas V.O and Sharma R.S, {\it Astrophysics and Space Science} {\bf356(2)} (2015) 173.
\bibitem[\protect \citeauthoryear{Thomas and Pandya}{2015}]{Pandya15b} Thomas V.O and D.M Pandya, {\it Astrophysics and Space Science} {\bf360} (2015) 59.
\bibitem[\protect \citeauthoryear{Chan {\em et al}}{1993}]{Chan93} Chan R., Herrera L., Santos N.O., {\it Mon.Not.R.Astron Soc.} {\bf265} (1993) 533.
\bibitem[\protect \citeauthoryear{Sharma {\em et al}}{2001}]{Sharma01} Sharma R. {\em et al} {\it Gen. Relativ. Gravit} {\bf33} (2001) 999.
\bibitem[\protect \citeauthoryear{Singh {\em et al}}{2016}]{SinghKSH16} Singh K.N, Bhar P. and Pant N.  {\it Astrophysics and Space Science} {\bf78} (2016) 339.
\bibitem[\protect \citeauthoryear{Ratanpal {\em et al}}{2016}]{Ratanpal16} Ratanpal B.S Thomas V.O and Pandya D.M, {\it Astrophysics and Space Science } {\bf45} (2016) 2016.
\bibitem[\protect \citeauthoryear{Bhar}{2015a}]{Bhar15a} Bhar P., {\it \apss} {\bf356} (2015) 309.
\bibitem[\protect \citeauthoryear{Thomas and Pandya}{2017}]{Pandya17} Thomas V.O and Pandya D.M, {\it Eur. Phys. J. A} {\bf53(6)} (2017) 123.
\bibitem[\protect \citeauthoryear{Singh and Pant}{2015}]{NewtonSingh15a} Singh K. N. and Pant N., {\it \apss} {\bf358} (2015) 44.
 \bibitem[\protect \citeauthoryear{L.Herrera}{1992}]{Herrera92} Herrera L., {\it Phys. Lett. A.} {\bf165} (1992) 206.
\bibitem[\protect \citeauthoryear{Karmarkar}{1948}]{Karmarkar48} Karmarkar K. R., {\it Proc. Indian Acad. Sci.} {\bf27} (1948) 56.
\bibitem[\protect \citeauthoryear{Bhar {\em et al}}{2016}]{Bhar16} Bhar P., Maurya S. K. Gupta Y. K. and Tuhina M., {\it Eur. Phys. J. A} {\bf 52} (2016) 312.
\bibitem[\protect \citeauthoryear{Singh and Pant}{2016}]{Singh16} Singh Ksh. Newton and Pant Niraj, {\it \apss} {\bf361} (2016) 177.
\bibitem[\protect \citeauthoryear{Singh {\em et al}}{2017}]{Singh17} Singh Ksh. Newton, Murad M. H. and Pant Niraj, {\it Eur. Phys. J. A} {\bf53} (2017) 21.
\bibitem[\protect \citeauthoryear{Özel and Güver}{2009}]{ozel09} F.Özel and T Güver, {\it Apj} {\bf693} (2009) 1775.
\bibitem[\protect \citeauthoryear{Gangopadhyay {\em et al}}{2013}]{Gangopadhyay13} Gangopadhyay T., Ray S., Li X-D., Dey J. and Dey M., {\it Mon. Not. R. Astron. Soc.} {\bf431} (2013) 3216.
\bibitem[\protect \citeauthoryear{Buchdahl}{1979}]{Buchdahl79} Buchdahl H. A., {\it Acta Phys. Pol.} {\bf10} (1979) 673.
\bibitem[\protect \citeauthoryear{Bayin}{1982}]{Bayin82} Bayin S.S, {\it Phys. Rev. D.} {\bf26} (1982) 6.
\bibitem[\protect \citeauthoryear{Kuchowicz}{1972}]{Kuchowicz72} Kuchowicz B., {\it Phys. Lett. A} {\bf38} (1972) 369.{\url{doi:10.1016/03759601(72)90164-8}}.
\bibitem[\protect \citeauthoryear{Murad and Fatema}{2015}]{Murad15} Murad M. H. and Fatema S., {\it Eur. Phys. J. C} {\bf75} (2015) 533.
\bibitem[\protect \citeauthoryear{Knutsen}{1987}]{Knutsen87} Knutsen H., {\it \apss} {\bf149} (1987) 38.
\bibitem[\protect \citeauthoryear{H.Hernández {\em et al}}{1999}]{Hernandez99} H.Hernández, L.A.Núñez, U.Percoco {\it Class. Quantum Grav.} {\bf16} (1999) 871.
\bibitem[\protect \citeauthoryear{T.Singh {\em et al}}{1995}]{Singh95} T.Singh, G.P.Singh, A.M.Helmi {\it II Nuovo Cim.B} {\bf110} (1995) 387.
\bibitem[\protect \citeauthoryear{K.Dev {\em et al}}{2003}]{DevGleiser03} K.Dev, M.Gleiser {\it Gen. Relativ. Gravit.} {\bf35} (2003) 1435.
\bibitem[\protect \citeauthoryear{M.Gleiser {\em et al}}{2004}]{DevGleiser04} M.Gleiser, K.Dev {\it Int. J. Mod. Phys. D} {\bf13} (2004) 1389.
\bibitem[\protect \citeauthoryear{T.Harko {\em et al}}{2000}]{HarkoMak00} T.Harko, M.K.Mak {\it J. Math. Phys.} {\bf41} (2000) 4752.
\bibitem[\protect \citeauthoryear{L.K.Patel{\em et al}}{1995}]{PatelMehta95}L.K.Patel, N.P.Mehta {\it Aust. J. Phys.} {\bf48} (1995) 635.
\bibitem[\protect \citeauthoryear{K.Lake {\em et al}}{2004}]{Lake04} K.Lake {\it Phys. Rev. Lett.} {\bf92} (2004) 051101.
\bibitem[\protect \citeauthoryear{C.G.Böhmer {\em et al}}{2006}]{BohmerHarko06} C.G.Böhmer, T.Harko {\it Class. Quantum Gravity} {\bf23} (2006) 6479.
\bibitem[\protect \citeauthoryear{C.G.Böhmer {\em et al}}{2007}]{BohmerHarko07} C.G.Böhmer, T.Harko {\it Mon. Not. R. Astron. Soc.} {\bf379} (2007) 393.
\bibitem[\protect \citeauthoryear{M.Esculpi {\em et al}}{2007}]{Esculpi07} M.Esculpi, M.Malaver, E.Aloma {\it Gen. Relativ. Gravit.} {\bf39} (2007) 633.
\bibitem[\protect \citeauthoryear{G.Khadekar {\em et al}}{2007}]{KhadekarTade07} G.Khadekar, S.Tade {\it Astrophys. Space. Sci.} {\bf310} (2007) 41.
\bibitem[\protect \citeauthoryear{S.Karmakar {\em et al}}{2007}]{Karmakar07} S.Karmakar, S.Mukherjee, R.Sharma, S.D.Maharaj {\it CPramana J.Phys.} {\bf68} (2007) 881.
\bibitem[\protect \citeauthoryear{H.Abreu {\em et al}}{2007}]{Abreu07} H.Abreu, H.Hernández, L.A.Núez{\it Class. Quantum Grav.} {\bf24} (2007) 4631.
\bibitem[\protect \citeauthoryear{B.V.Ivanov {\em et al}}{2010}]{Ivanov10} B.V. Ivanov. {\it Int. J. Mod. Phys. A} {\bf25} (2010) 3975.
\bibitem[\protect \citeauthoryear{L.Herrera {\em et al}}{2008a}]{Herrera08}L.Herrera, N.O.Santos, A.Wang {\it Phys. Rev. D} {\bf78} (2008) 084026.
\bibitem[\protect \citeauthoryear{M.K.Mak {\em et al}}{2003}]{MakHarko03} M.K.Mak,T.Harko {\it Proc. R. Soc. Lond.} {\bf A459} (2003) 393.
\bibitem[\protect \citeauthoryear{R.Sharma {\em et al}}{2002}]{SharmaMukherjee02} R.Sharma, S.Mukherjee {\it Mod. Phys. Lett. A} {\bf17} (2002) 2535.
\bibitem[\protect \citeauthoryear{T.Harko {\em et al}}{2002}]{HarkoMak02} T.Harko, M.K.Mak {\it Annalen der Phys.} {\bf11} (2002) 3.
\bibitem[\protect \citeauthoryear{L.Herrera {\em et al}}{2008b}]{HerreraOspino08} L.Herrera, J.Ospino, A.di Prisco {\it Phys.Rev.D} {\bf77} (2008) 027502.
\bibitem[\protect \citeauthoryear{K.D.Krori {\em et al}}{1984}]{Krori84} K.D.Krori, P.Borgohain, R.Devi {\it Can.J.Phys.} {\bf62} (1984) 239.
\bibitem[\protect \citeauthoryear{H. Bondi {\em et al}}{1993}]{Bondi93} H. Bondi {\it Mon. Not.R.Astron.Soc.} {\bf262} (1993) 1088.
\bibitem[\protect \citeauthoryear{H. Bondi {\em et al}}{1999}]{Bondi99} H. Bondi {\it Mon. Not.R.Astron.Soc.} {\bf302} (1999) 337.
\bibitem[\protect \citeauthoryear{W. Barreto {\em et al}}{1993}]{Barreto93} W. Barreto {\it Astrophys.Space.Sci.} {\bf201} (1993) 191.
\bibitem[\protect \citeauthoryear{W. Barreto {\em et al}}{2007}]{Barreto07} W. Barreto, B. Rodríguez, L. Rosales, O. Serrano {\it Gen. Relativ. Gravit.} {\bf39} (2007) 23.
\bibitem[\protect \citeauthoryear{ A.A. Coley {\em et al}}{1994}]{Coley94} A.A. Coley, B.O.J. Tupper {\it Class. Quantum Gravity} {\bf11} (1994) 2553.
\bibitem[\protect \citeauthoryear{J. Martínez {\em et al}}{1994}]{Martinez94}J. Martínez, D. Pavón, L.A. Nuñez {\it Mon. Not. R. Astron. Soc.} {\bf271} (1994) 463.
\end{thebibliography}
\end{document}